\documentclass{elsart4-1}

\usepackage{graphicx}
\usepackage{epstopdf}
\usepackage{amssymb}
\usepackage{amsmath}

\usepackage[english,francais]{babel}

\newtheorem{e-proposition}[theorem]{Proposition}

\newtheorem{e-definition}[theorem]{Definition\rm}

\renewcommand{\theequation}{\arabic{equation}}
\def\og{\leavevmode\raise.3ex\hbox{$\scriptscriptstyle\langle\!\langle$~}}
\def\fg{\leavevmode\raise.3ex\hbox{~$\!\scriptscriptstyle\,\rangle\!\rangle$}}

\begin{document}
% Select a primary header Physics or Astrophysics
% You can place after the header (classification), if you know it.

%\sloppy

\centerline{Astrophysics}
\begin{frontmatter}


% \title{The Ankle}%\thanksref{label1}}
% \author{Name\thanksref{label2}}

\selectlanguage{english}
\title{Observations of supernova remnants and pulsar wind nebulae at gamma-ray energies}

% If all authors are at the same address, the [label1] can be suppressed

\selectlanguage{english}
\author[a]{John W. Hewitt}
\ead{john.w.hewitt@nasa.gov}
\author[b]{\and Marianne Lemoine-Goumard}
\ead{lemoine@cenbg.in2p3.fr}
\address[a]{NASA/GSFC\\ Mail Code 661, Building 34, Greenbelt, MD 20771 USA}
\address[b]{CENBG\\ Chemin du Solarium, CS 10120, 33170 GRADIGNAN CEDEX, France}

\medskip

\begin{abstract}
{\bf Acceleration of cosmic rays at supernova remnant and pulsar wind nebulae shocks: constraints from gamma-ray observations\\ }
In the past few years, gamma-ray astronomy has entered a golden age thanks to two major breakthroughs: Cherenkov telescopes on the ground and the Large Area Telescope (LAT) onboard the \emph{Fermi} satellite. The sample of supernova remnants (SNRs) detected at gamma-ray energies is now much larger: it goes from evolved supernova remnants interacting with molecular clouds up to young shell-type supernova remnants and historical supernova remnants. Studies of SNRs are of great interest, as these analyses are directly linked to the long standing issue of the origin of the Galactic cosmic rays. In this context, pulsar wind nebulae (PWNe) need also to be considered since they evolve in conjunction with SNRs. As a result, they frequently complicate interpretation of the gamma-ray emission seen from SNRs and they could also contribute directly to the local cosmic ray spectrum, particularly the leptonic component. This paper reviews the current results and thinking on SNRs and PWNe and their connection to cosmic ray production.

\vskip 0.5\baselineskip

\selectlanguage{francais}
\noindent{\bf R\'esum\'e}
\vskip 0.5\baselineskip
\noindent
{\bf Acc\'el\'eration des rayons cosmiques par les vestiges de supernovae et les n\'ebuleuses de pulsars: contraintes fournies par les observations gamma\\ }
Au cours des derni\`eres ann\'ees, l'astronomie gamma est entr\'ee dans un \^age d'or gr\^ace \`a deux avanc\'ees majeures: les t\'elescopes Tcherenkov au sol et le Large Area Telescope (LAT) \`a bord du satellite Fermi. L'\'echantillon des restes de supernova (SNR) d\'etect\'es en rayons gamma de haute \'energie est maintenant beaucoup plus vaste: il va des vestiges de supernova \'evolu\'es en interaction avec des nuages mol\'eculaires jusqu'aux jeunes SNR en coquille et aux SNR historiques. Les \'etudes des SNR sont d'un grand int\'er\^et car ces analyses sont directement li\'ees \`a la question de l'origine des rayons cosmiques galactiques. Dans ce contexte, les n\'ebuleuses de pulsars (PWN) doivent \'egalement \^etre prises en compte car elles \'evoluent en conjonction avec les SNRs. En cons\'equence, elles compliquent souvent l'interpr\'etation de l'\'emission gamma en provenance des SNR et pourraient aussi contribuer directement au spectre local de rayons cosmiques, en particulier \`a sa composante leptonique. Cet article passe en revue les r\'esultats et r\'eflexions actuels concernant les SNR et les PWN ainsi que leur connexion \`a la production des rayons cosmiques.

\vskip 0.5\baselineskip

\noindent{\small{\it Keywords~:} Cosmic rays; Supernova remnants; Pulsar wind nebulae \vskip 0.5\baselineskip
\noindent{\small{\it Mots-cl\'es~:} Rayons cosmiques~; vestiges de supernovae~; N\'ebuleuses de pulsar}}

\end{abstract}
\end{frontmatter}

% now the Version française abrégée, if it exists
%\selectlanguage{francais}
%\section*{Version fran\c{c}aise abr\'eg\'ee}
% Text of your Version française abrégée here

\selectlanguage{english}

%%%%%%%%%%%%%%%%%%%%%%%%%%%%%%%%%%%%%%%%%%%%%%%%%%%%%%%%%
\section{Cosmic-ray acceleration at supernova remnant shocks}
\label{sec:cosmic}

Supernova remnants (SNRs) have been considered as the sources of Galactic cosmic rays (CRs) since the 1930's. Support for this paradigm comes from observed CR abundances that match those of massive star forming regions, non-thermal emission from SNRs indicative of relativistic electron (and possibly proton) acceleration, and a theoretical mechanism that explains how particles gain relativistic energies in the strong shocks of SNRs by first order Fermi diffusive shock acceleration (DSA) \cite{dsa}. The high efficiency of DSA has been demonstrated by non-linear models \cite{Bell1987,Jones1991}, though the mechanisms by which particles are injected and confined within the shock vicinity are still debated. It is also unclear how particle confinement and escape is regulated within SNRs, which is necessary for CRs to reach energies of a few times 10$^{15}$ eV that are attributable to Galactic sources. Finally, it is important to study the diffusion parameters of accelerated particles as they leave the dense star forming regions where they are thought to arise, to understand the spectrum and isotropy of CRs that permeate the galaxy.
\\

The recent development of gamma-ray observatories has led to sufficient sensitivity and spatial resolution to directly study SNRs as CR sources. Ground-based Cherenkov telescopes (e.g. \textit{HESS, MAGIC, VERITAS}) and the \textit{Fermi Gamma-ray Space Telescope} are able to spatially resolve SNRs at TeV and GeV energies, respectively. To date, dozens of SNRs have been detected varying in age, progenitor type, evolutionary stage and density of their surrounding environment. These observations show a diversity in the observed luminosities and spectra of SNRs that indicate a wide range of physical conditions that give rise to gamma rays. Either electrons or protons may dominate the radiation mechanism, and the large energy coverage of \textit{Fermi} (from 0.03 to $>$300 GeV) is important in distinguishing between leptonic (inverse-Comtpon (IC) or bremsstrahlung) and hadronic ($\pi^0$ decay) processes. Observations of young SNRs (section \ref{sec:young}) help to establish the earliest periods of DSA and possible proton acceleration to very high energies. Studies of more numerous shell-type (section \ref{sec:shell}) and middle-age SNRs (section \ref{sec:middle}) provide tests of both the acceleration and propagation of CRs from their acceleration sites.

\subsection{Diffusive shock acceleration theory}

In strong shocks, particles in either the preshock or postshock fluid have scattering centers such that they experience only approaching collisions. This allows energy gains of first-order as particles cross the shock front, with a spectrum that depends only on the shock compression ratio as explained in \cite{dsa}. This formulation of DSA or first-order Fermi acceleration was proposed by several groups to explain radio-emitting electrons discovered in SNRs \cite{Axford1977,Krymskii1977,Bell1978,BO1978}. However, the GeV energies of radio synchrotron electrons are far less than needed to explain PeV energy CRs. More complex models have evolved to include heating of the upstream precursor, the pressure produced by accelerated particles, injection of electrons and protons, and different prescriptions for magnetic field amplification needed to confine the highest-energy particles e.g. \cite{Berezhko1999}. Predictions of such models includes CR source spectra whose spectral indices are greater ("soft" spectra) or lower ("hard" spectra) than 2.0 and can have spectral curvature. For an in-depth review of advances in shock acceleration theory, see \cite{dsa} in this volume.

\subsection{Supernova remnant evolution}

Given the non-linearity of DSA, the evolution of SNRs has implications for when and at what efficiency CRs are produced. The key parameters governing the evolution of a supernova (SN) explosion are the initial energy, progenitor system, and surrounding interstellar medium (ISM), which are all tied together by the late-time evolution of massive stars. The progenitor system can be either a binary system (two objects orbiting one another) composed of a white dwarf and a companion star for Type Ia SNe, or a massive star (M $>$ 8 $M_{\odot}$) for type II SNe (also known as core-collapse SNe), or even massive stars that have lost their Hydrogen-envelopes prior to explosion (Wolf-Rayet stars, $>$ 25 $M_{\odot}$) for Type Ib/c SNe. 
%
% the reader should be briefly reminded about the two main types of supernovae (core-collapse and thermonuclear or SN Ia) before considering the evolution.
%
The evolution of a SNR is then typically characterized by three phases: 1. an ejecta-dominated phase when the swept-up mass is much less than the ejected mass and does not slow shock expansion, 2. when the swept-up mass is much larger than the ejecta mass and the shock expansion is characterized by adiabatic expansion often called the Sedov-Taylor phase, 3. a radiative phase where recombining gas quickly cools the swept-up gas forming a dense shell \cite{Woltjer1970}. Toward the end of this radiative phase, the dense shell of the SNR merges with and becomes indistinguishable from the ISM \cite{Cioffi1988}. Simple evolutionary models with prescriptions for shock acceleration reproduce the expectation that $\sim 10$\% the SN explosion kinetic energy is transferred to CRs, which is required if SNRs are the primary source of Galactic CRs \cite{Drury1989}.
% Discuss Acceleration at the reverse shock?
Depending on the density profile of the surrounding gas, SNRs may transition between these phases at very different times \cite{Dwarkadas1998}. The result of models that include evolution in complex density profiles indicate that young core-collapse SNRs will have decreasing hadronic gamma-ray emission, while Type Ia SNe are expected to increase their gamma-ray emission with time~\cite{Dwarkadas2013}. However, when the SNR shock evolves into a dense cloud or wind-blown bubble, the ambient density encountered by the shock rapidly increases, enhancing the target density for hadronic gamma rays while reducing shock velocities to no more than a few hundred km s$^{-1}$ \cite{Chevalier1999}. It has been proposed that such systems may still accelerate CRs early during the interaction via reflected shocks produced when the shock wave hits the clouds \cite{Inoue2012}.

\subsection{Evolution of composite supernova remnants}

\begin{figure}[t]
 \centering
  \includegraphics[height=0.3\textheight]{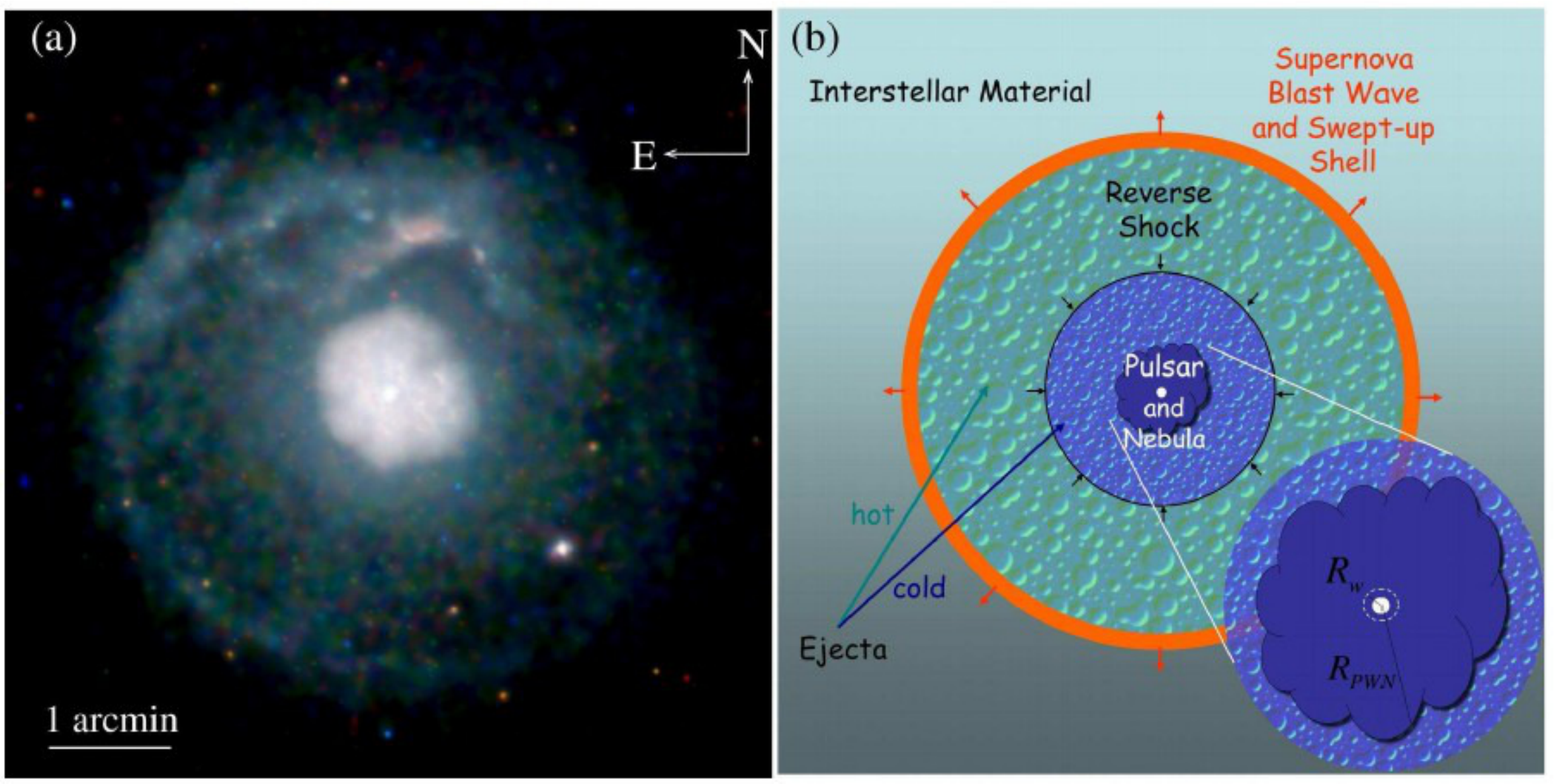}
  \caption{Left panel: Chandra X-ray image of the composite SNR G21.5-0.9. The spherical SN shell is faintly glowing in X-rays; the center of the remnant is filled by a PWN of highly relativistic electrons with strong X-ray emission. Right panel: Conceptual diagram of a PWN within an SNR, illustrating the central pulsar and nebula, surrounding cold ejecta, shock-heated ejecta,
and surrounding medium. This Figure is taken from \cite{pwn_review}.
}
  \label{fig:pwn}
\end{figure}

Some SNe leave behind a rapidly spinning neutron star. These pulsars dissipate their rotational energy via a relativistic wind of electrons and positrons, which is rapidly decelerated at an interaction point with the surrounding dense medium called the termination shock \cite{pwn_review,pulsar}. The pulsar wind nebula (PWN) formed in the interior of the SNR expands outward and may encounter the inward reverse shock of the SNR as illustrated in Figure~\ref{fig:pwn}. These systems are typified by a radio synchrotron shell for the SNR and a flat-spectrum central PWN, and therefore dubbed composite SNRs. Relativistic electrons produce IC gamma rays at GeV and TeV energies. Additional gamma-ray emission from the pulsar magnetosphere and surrounding SNR shock front can also be present, limiting the identification of the primary source of gamma rays. At various points during the evolution of the system, different regions may produce the brightest gamma-ray emission. Models of the evolution of PWNe and SNRs show a diversity of possible observed parameters \cite{Gelfand2009,Martin2012}. While shocks within PWN are clearly capable of accelerating leptons, it is not clear if protons are also accelerated, or how particles diffuse from the nebula and enter the ISM (see section \ref{sec:positron}).

%%%%%%%%%%%%%%%%%%%%%%%%%%%%%%%%%%%%%%%%%%%%%%%%%%%%%%%%%
\section{Young supernova remnants as ideal proton accelerators viewed at gamma-ray energies}
\label{sec:young}

\subsection{SN 1987A}

Located in the Large Magellanic Cloud at only 50 kpc from Earth, SN 1987A is the closest observed with modern telescopes. After evolving for nearly three decades, the radio flux has been steadily increasing, suggesting increasing acceleration, albeit with a steep spectrum (F$_\nu \propto \nu^\alpha$ where $\alpha$ = $-0.8\pm0.1$, where F$_\nu$ is the energy received per unit area, unit time and unit frequency) with no evidence of a spectral break \cite{2014ApJ...782L...2I}. This is inconsistent with expectations from diffusive shock acceleration, and perhaps indicative that the shock is modified by pressure from accelerated particles \cite{Ellison2000} or by an amplified magnetic field \cite{Reville2013}. A flat-spectrum radio excess appears at the center of the remnant, and may be attributed to a newly formed PWN \cite{2014ApJ...796...82Z}. Even with a deep exposure of 210 hours, significant emission from SN 1987A is not detected by H.E.S.S. \cite{hess_lmc}. The 99\% flux limit above 1 TeV is $5.6\times10^{-14}$ cm$^{-2}$ s$^{-1}$ assuming a photon index of 2, which translates into an upper limit for the $\gamma$-ray luminosity of $2.2\times10^{27}$ W ($2.2\times10^{34}$ erg s$^{-1}$). Multiwavelength studies of SN 1987A suggest that the shock at the current epoch has reached the dense equatorial ring made by the progenitor system for which densities of $10^3$ to $10^4$ cm$^{-3}$ have been found \cite{dens_sn1987a}. This implies that less than 1\% of the explosion energy of $10^{44}$~J ($10^{51}$ erg) is carried by accelerated CR nuclei. Since it is not clear when, during the transition from SN to SNR, shock acceleration becomes efficient, these gamma-ray observations of the youngest nearby SN are a strong help to constrain theories in the future.

\subsection{Cassiopeia A and Tycho}

Two historical SNRs have been detected both at GeV and TeV energies: Cassiopeia A (Cas~A) \cite{fermi_casa,magic_casa,veritas_casa} and Tycho \cite{fermi_tycho,veritas_tycho}. Cas~A is the remnant of SN 1680. It is the brightest radio source in our Galaxy and its overall brightness across the electromagnetic spectrum makes it a unique laboratory for studying high-energy phenomena in SNRs. A simple multiwavelength modeling of Cas~A strongly favors the hadronic scenarios and implies that the total content of CRs accelerated in Cas~A is $\sim$4$\times$10$^{42}$~J (4$\times$10$^{49}$~erg), and the magnetic field amplified at the shock can be constrained as B $\approx$ 12~nT (0.12~mG). Even though Cas A is considered to have entered the Sedov phase, the total amount of CRs accelerated in the remnant constitutes only a minor fraction ($\sim2$\%) of the total kinetic energy of the SN \cite{yuan_casA}, which is well below the $\sim 10$\% commonly used to maintain the CR energy density in the Galaxy.\\ 

Tycho's SNR (SN 1572) is classified as a Type Ia (thermonuclear explosion of a white dwarf) based on observations of the light-echo spectrum. Thanks to the large amount of data available at various wave bands, this remnant can be considered one of the most promising object where to test the shock acceleration theory and hence the CR -- SNR connection. First, using the precise radio and X-ray observations of this SNR, \cite{morlino} have shown that the magnetic field at the shock has to be $>$ 20~nT (0.2~mG) to reproduce the data. Then, using  multiwavelength data, especially the above mentioned GeV and TeV detections, they could infer that the gamma-ray emission detected from Tycho cannot be of leptonic origin, but has to be due to accelerated protons (this result is consistent with another modeling proposed in \cite{fermi_tycho}). These protons are accelerated up to energies as large as $\sim$500 TeV, with a total energy converted into CRs estimated to be about 12\% of the forward shock bulk kinetic energy. This is fully consistent with acceleration of Galactic CRs in SNRs.\\

\subsection{Other young Galactic supernova remnants}

Two other Galactic SNRs are known to have similarly young ages but have not yet been detected in gamma rays. Kepler's SNR (SN 1604) is a type Ia or II-L SN \cite{chiotellis12} with detected synchrotron X-rays. Detailed modeling predicts detectable gamma rays for existing observatories in both GeV and TeV regimes. However, the non-detection by H.E.S.S. may be explained if the distance is at least 6.4 kpc, or the magnetic field behind the shock is greater than 5.2 nT (52 $\mu$G) \cite{hess_kepler}. Hydrodynamical models suggest a distance of at least 7 kpc \cite{patnaude12}. SNR G1.9+0.3 has an age of 181$\pm$25 years estimated from proper motions, making it the youngest known Galactic SNR \cite{2014SerAJ.189...41D}. Interestingly, the parent SN was not visible since it has been obscured by the dense gas and dust of the Galactic Center. Synchrotron emission from G1.9+0.3 from radio and X-rays shows electrons with a power-law index of $2.27$ and a roll-off frequency (peak frequency in the spectral energy distribution, SED, emitted by electrons with energy $E_{\rm max}$) of ($3.07 \pm 0.18$)$\times$10$^{17}$ Hz \cite{2015ApJ...798...98Z}. Knowing that the roll-off frequency $\nu_{\rm rolloff}$ of the spectrum is related to the the maximum electron energy, $E_{\rm max}$, following the relation 
% the "starred" equation environments produce no equation numbers
\begin{align*} 
\nu_{\rm rolloff}={\rm 5}\times10^{15} \; {\rm Hz} \; \biggl(\frac{\it B}{1 \; nT}\biggr) \; \biggl(\frac{{\it E}_{\rm max}}{10 \; {\rm TeV}}\biggr)^2,
\end{align*}
%$\nu_{\rm rolloff}={\rm 5}\times10^{15} \; {\rm Hz} \; \biggl(\frac{\it B}{1 \; nT}\biggr) \; \biggl(\frac{{\it E}_{\rm max}}{10 \; {\rm TeV}}\biggr)^2$,
we can deduce that for a magnetic field of about 1 nT (10 $\mu$G) the maximum energy of accelerated particles is about 80 TeV. If the roll-off in the spectrum results from cooling and is not age limited, higher energies may be reached by nuclei. H.E.S.S. observations set an upper limit on the photon flux from this remnant of 5.6$\times$10$^{-13}$ cm$^{-2}$ s$^{-1}$ above 0.26 TeV, which in a one-zone leptonic scenario places a lower limit on the interior magnetic field of 1.2 nT (12 $\mu$G) \cite{hess_g1.9}.
Future gamma-ray detection of either remnant will help to diversify the sample of young, active accelerators that can be studied. There is also potential for gamma-ray surveys to identify new young Galactic SNRs. Radio surveys which are commonly used to identify SNRs are biased against small angular diameter SNRs, as would be the case for a young remnant on the other side of the Galaxy. Future surveys with CTA could reach sufficient sensitivity to take a census of the young Galactic SNRs \cite{cta_young}.

%%%%%%%%%%%%%%%%%%%%%%%%%%%%%%%%%%%%%%%%%%%%%%%%%%%%%%%%%
\section{Shell-type supernova remnants, efficient electron accelerators}
\label{sec:shell}

Five young SNRs with clear shell-type morphology resolved in VHE gamma rays have been detected by H.E.S.S.: RX J1713.7-3946 \cite{aharonian_rxj1,aharonian_rxj2}, RX J0852.04622 - also known as Vela Junior - \cite{aharonian_velajr1}, RCW 86 \cite{aharonian_rcw86}, SN 1006 \cite{acero_sn1006} and HESS J1731-347 \cite{acero_1731}. Three of them, RX~J1713.7-946 \cite{fermi_rxj}, Vela Junior \cite{fermi_velajr}, and recently RCW 86 \cite{yuan_rcw86}, have been detected by \emph{Fermi}-LAT allowing direct investigation of young shell-type SNRs as sources of CRs. The shell-type morphology of the gamma-ray emission, which is associated with the SN blast wave, provides convincing evidence that DSA is the mechanism producing the high energy particles radiating at gamma-ray energies. A critical issue for DSA, and for the origin of CRs, however, concerns the radiation mechanism responsible for the GeV$-$TeV emission: is it dominated by $\pi^0$-decay emission from nuclei or IC emission from leptons?

\begin{figure}[!t]
    \centering
    \includegraphics[height=0.3\textheight]{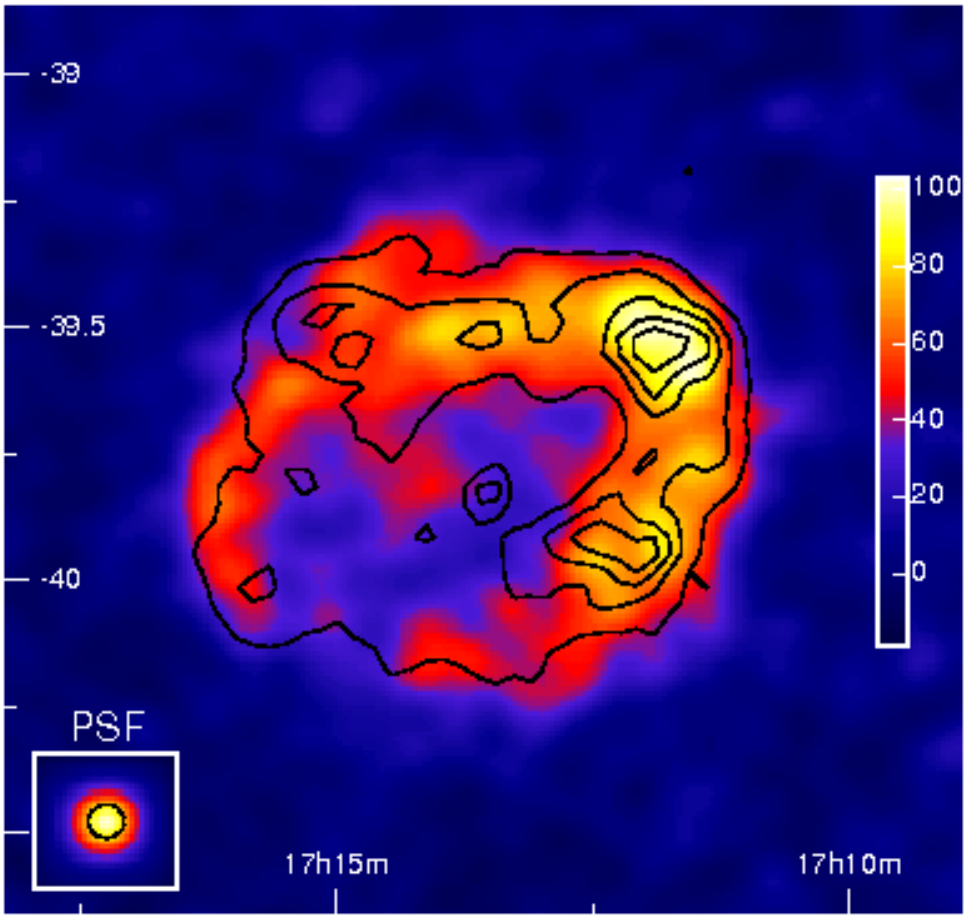}
     \includegraphics[height=0.3\textheight]{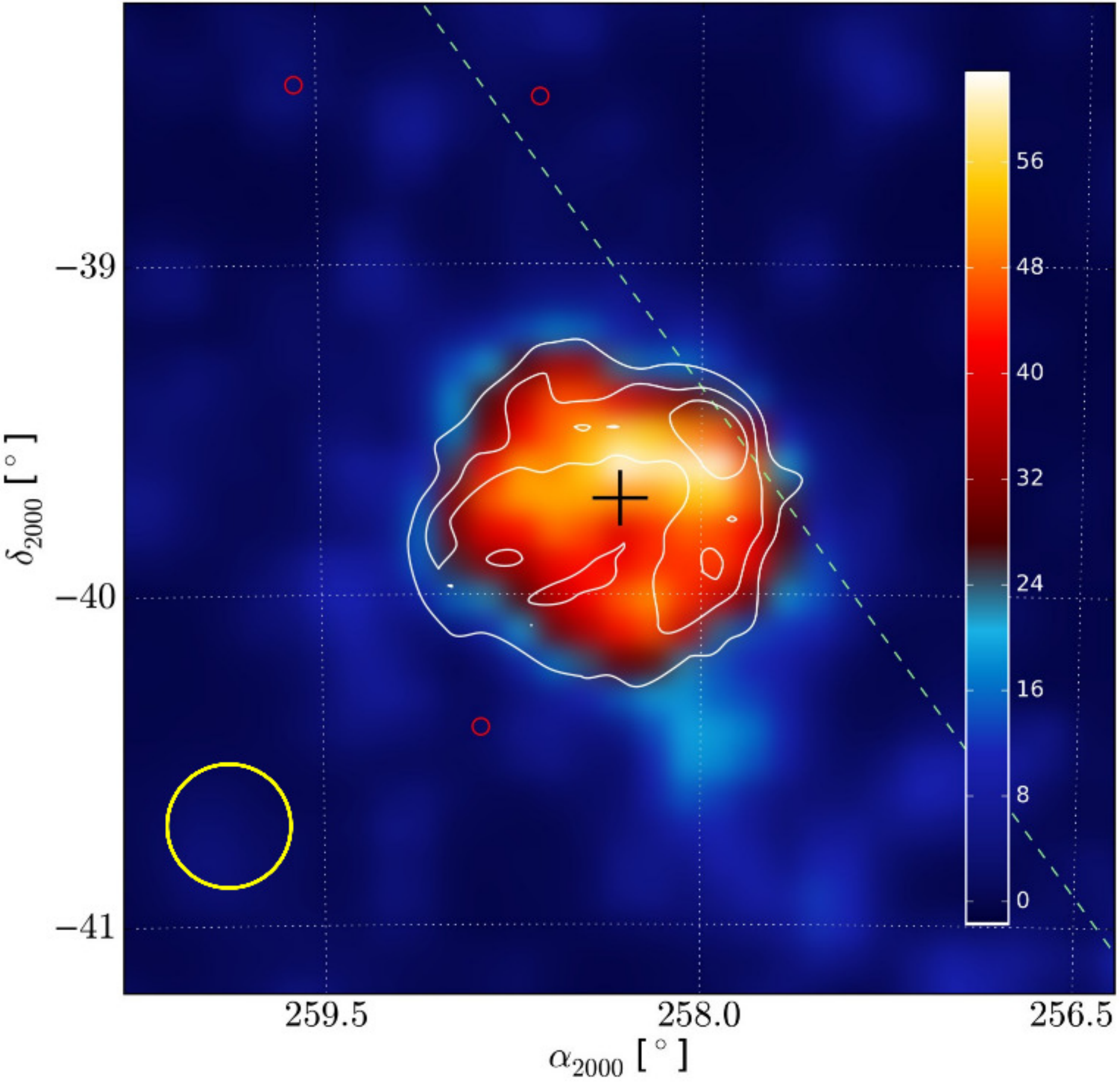}
    \caption{Left: Combined H.E.S.S. image of RX~1713.7$-$3946 from the 2004 and 2005 data from \cite{aharonian_rxj2}. The image is smoothed with a Gaussian of  2 arcmin and the linear colour scale is in units of excess counts per smoothing radius. X-ray contours from ASCA are drawn as black lines~(1-3~keV, from \cite{UchiyamaAsca}) for comparison. Right: \emph{Fermi}-LAT test statistic (TS) map of the region around RX~1713.7$-$3946 \cite{federici}. The linear colour scale is in units of TS which is an indicator of the significance of the detection of a point source at each point of the skymap (TS of 25 being close to a 5$\sigma$ detection). Shown in white are contours of gamma-ray excess counts based on H.E.S.S. observations, the TS levels are 25, 50, and 75. The insets in both figures represent the angular resolution (point-spread functions) for each analysis.}
    \label{fig:rxj}
  \end{figure}

\subsection{The famous case of RX J1713.7$-$3946}
RX J1713.7$-$3946 (also known as G347.3$-$0.5) is a young 'historical' remnant suggested to be associated with the appearance of a guest star in the constellation of Scorpius in AD393 by \cite{wang_rxj}. RX J1713.7$-$3946 is located in the Galactic plane (at a distance of 1kpc) and was discovered in soft X-rays in 1996 in the ROSAT all-sky survey \cite{rxj_rosat}. In X-rays RX J1713.7$-$3946 has a diameter of $1^{\circ}$, twice the size of the full moon. It is the first SNR for which TeV gamma-ray emission was clearly detected emerging from the shell \cite{aharonian_rxj1}. The gamma-ray emission as detected by H.E.S.S. and \emph{Fermi}-LAT closely matches the non-thermal X-ray emission as can be seen in Figure~\ref{fig:rxj}. The TeV spectrum of RX J1713.7$-$3946 is certainly the most precisely measured among the SNR shell class and shows significant emission up to $\sim$100 TeV, clearly demonstrating particle acceleration to beyond these energies in the shell of the SNR. However, the GeV spectrum measured with \emph{Fermi}-LAT is described by a very hard power-law with a photon index of $\Gamma = 1.5 \pm 0.1$. This hard GeV spectrum is in contradiction with most hadronic models published so far (e.g. \cite{berezhko}) and requires an unrealistically large density of the medium. Alternative and more complex hadronic scenarios have been recently introduced to fit the GeV-scale emission from RX J1713.7$-$3946, for instance the possibility of a shell of dense gas located a short distance upstream of the forward shock that would be illuminated by the runaway CRs that have escaped from the shock and by the CR precursor to the forward shock (see \cite{federici} for more details). In leptonic scenarios, the agreement with the expected IC spectrum is better but requires a very low magnetic field of $\sim$ 1~nT (10~$\mu$G) in comparison to the one measured in the thin filaments by X-ray observations. It is possible to reconcile a high magnetic field with the leptonic model if GeV gamma rays are radiated not only from the filamentary structures seen by Chandra, but also from other regions in the SNR where the magnetic field may be weaker. 

\begin{figure}[!t]
  \centering
  \includegraphics[height=0.3\textheight]{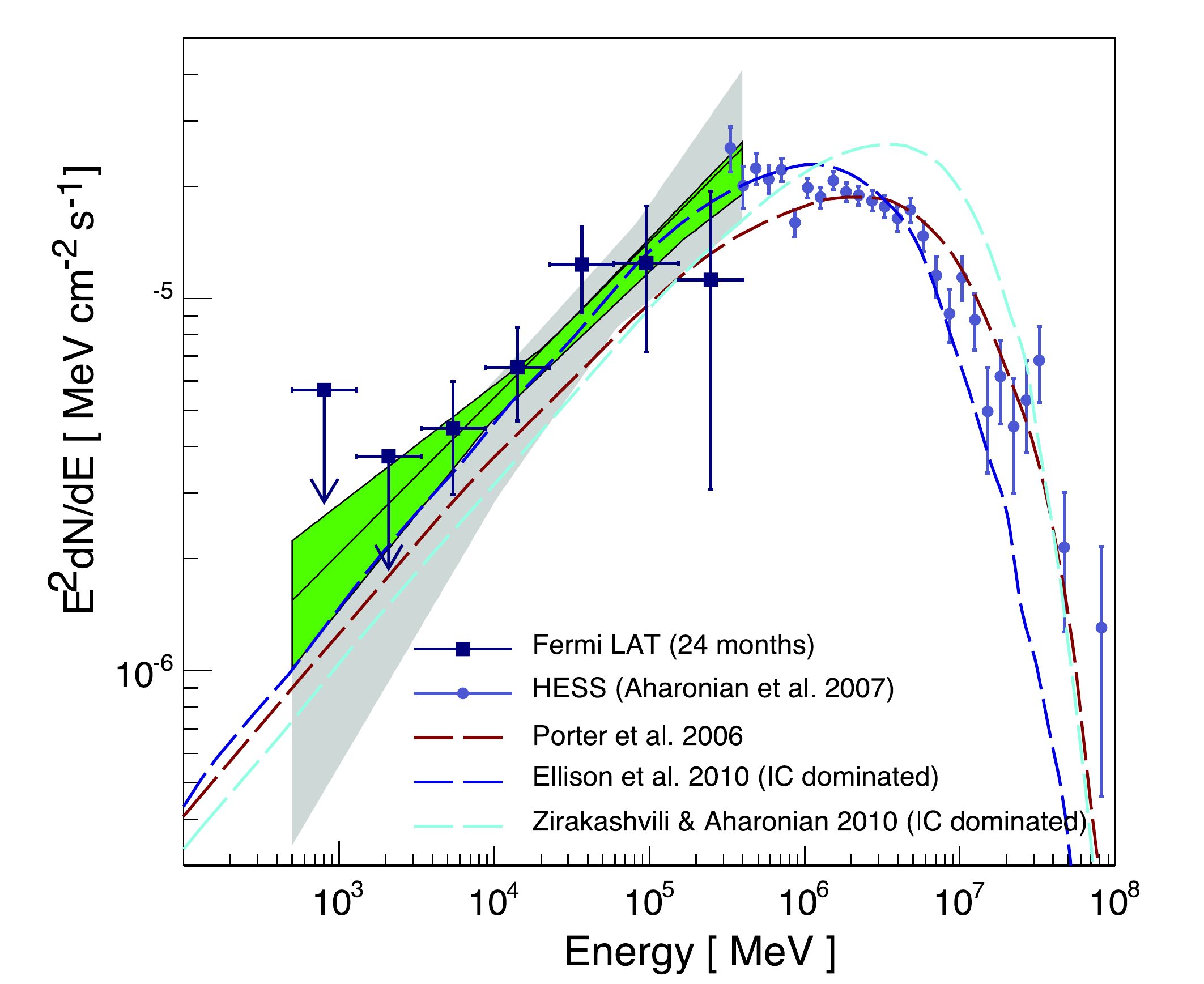}
  \includegraphics[height=0.3\textheight]{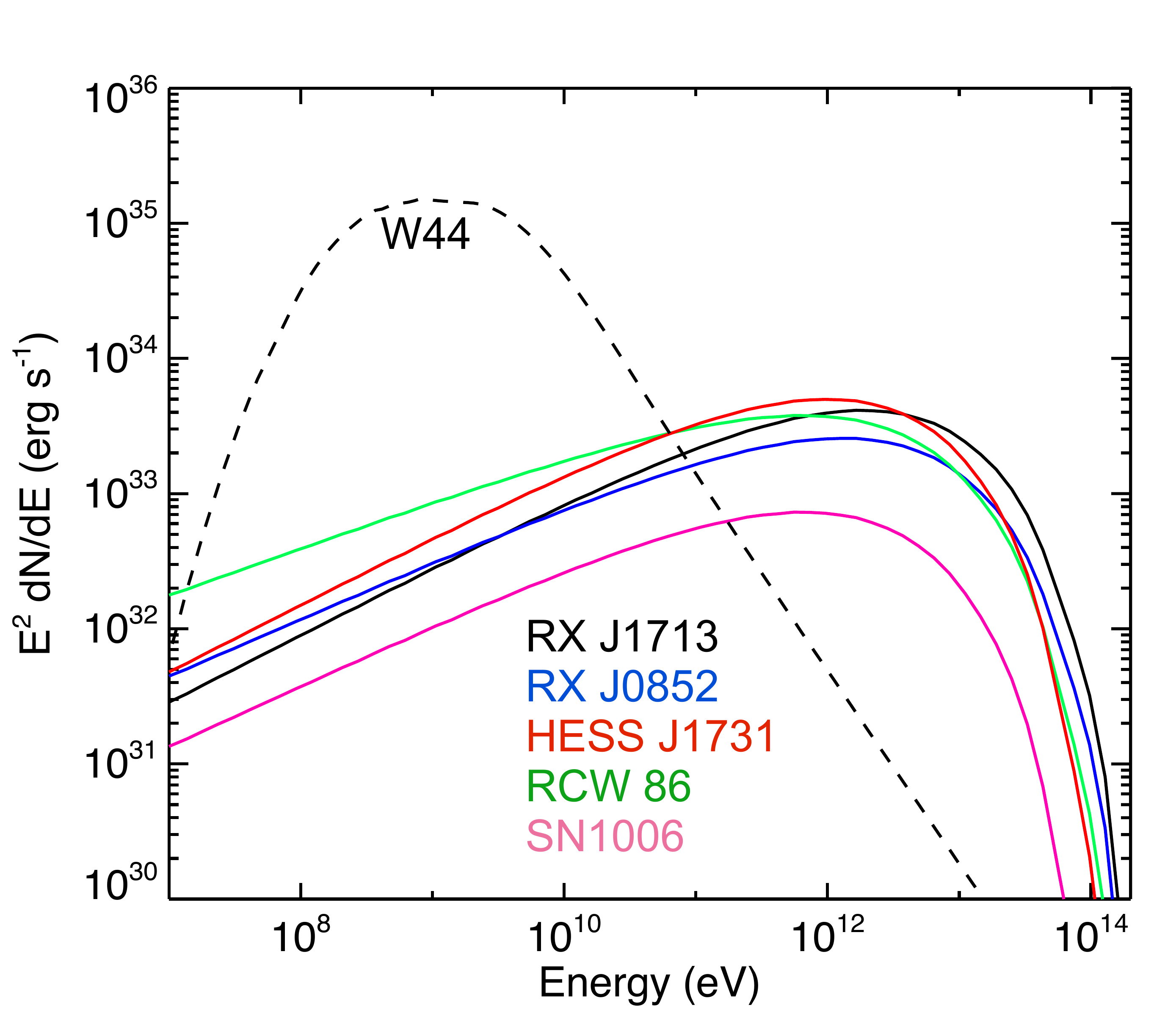}
  \caption{
  Left: Energy spectrum of RX J1713.7$-$3946 in gamma rays with different leptonic models generated to fit the TeV data points from the literature. From~\cite{fermi_rxj}.
  Right: Gamma-ray luminosity of the five shell-type SNRs detected at gamma-ray energies. For the sake of comparison the SED from the SNR W44 is shown by the dashed line. From~\cite{acero_shell}.}
  \label{fig:shell}
\end{figure}

\subsection{General properties of this supernova remnant population}
Interestingly, as can be seen in Figure \ref{fig:shell}, the five shell-type SNRs detected at gamma-ray energies show hard HE spectral indices ($1.4 < \Gamma < 1.8$) that exclude the standard hadronic test particle scenario\footnote{The test particle scenario neglects the retro-action of the CRs on the shock structure and produces a proton energy distribution which converts into a photon spectral index $\Gamma = 2$.}. All photon indices (except for RX J0852.0$-$4622) are compatible with a test particle leptonic dominated scenario where the electron slope of 2.0 translates into a photon spectral index of 1.5. In the case of RX J0852.0$-$4622, the slightly softer HE photon index ($1.85 \pm 0.06_{stat} \pm 0.19_{syst}$, \cite{fermi_velajr}) could be due to a deviation from the test particle case, a mix of hadronic and leptonic contributions or a possible contamination from the PWN seen around PSR J0855$-$4644 \cite{Acero2013} that is located right on the south-eastern part of the SNR shell. The similarity of hard photon spectral indices in this SNR sample tends to point towards a common leptonic dominated scenario for the HE and VHE gamma-ray emission. Still, there could be some smaller subregions (e.g. dense clumps) where the hadronic mechanism significantly contribute to the local gamma-ray emission.\\

\begin{figure}
\begin{center}
\includegraphics[width=0.8\linewidth]{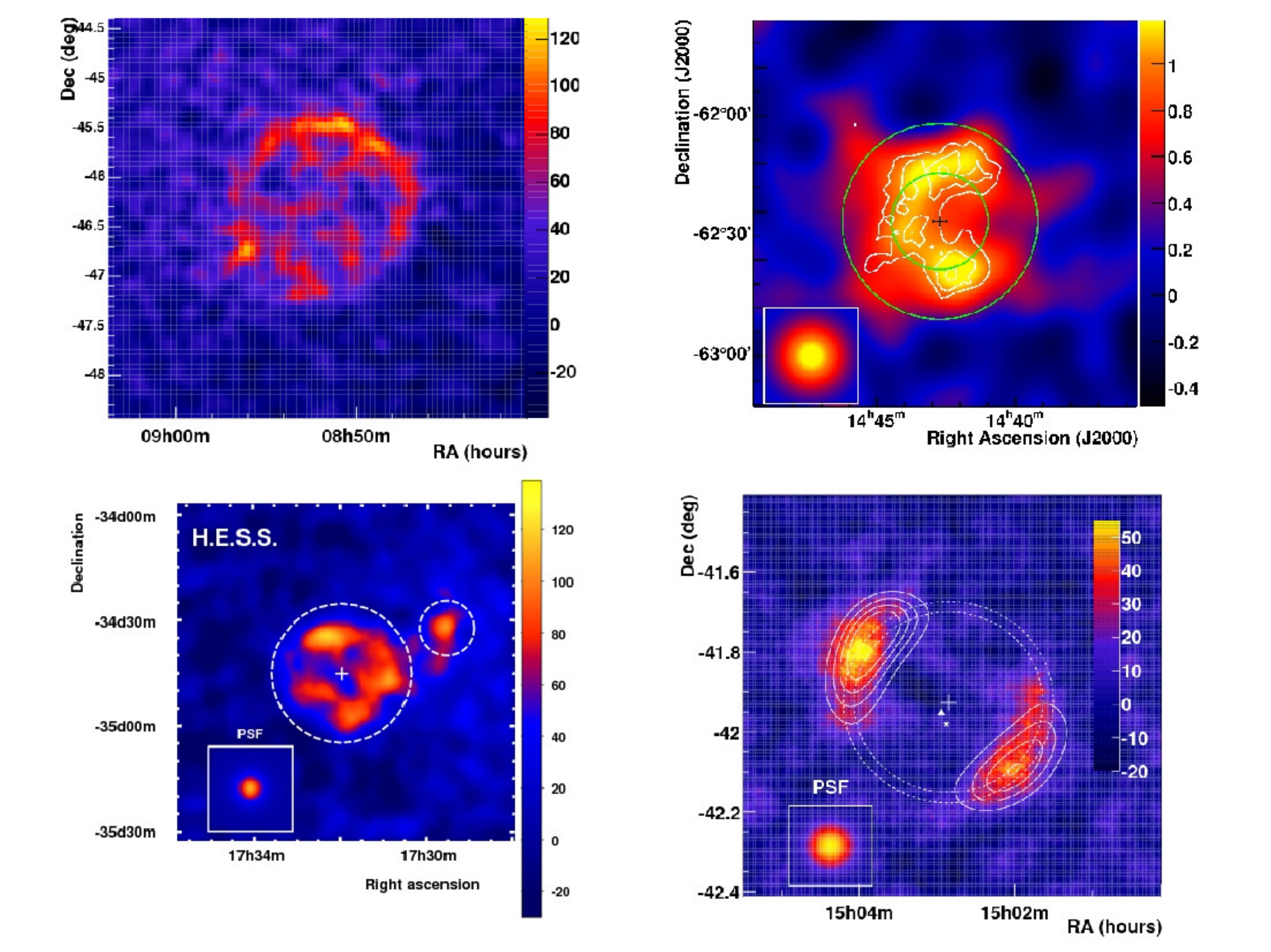}
\caption{H.E.S.S. $\gamma$-ray image of Vela Junior from \cite{aharonian_velajr1} (top left, linear colour scale is in units of excess counts per bin), RCW 86 \cite{aharonian_rcw86} (top right, linear color scale is in units of excess counts per arcmin$^{2}$), HESS J1731-347 \cite{acero_1731} (bottom left, linear colour scale is in units of excess counts per bin), and SN~1006 from \cite{acero_sn1006} (bottom right, linear colour scale is in units of excess counts per bin). }
\label{fig:snrs}
%% ==> sn1006_xmm_superposition.C 
\end{center}
\end{figure}

In addition to a similar HE spectral index, these five shell-type SNRs present a striking similarity in terms of peak luminosity and spectral shape. This contrast is highlighted when compared with the SNR W44 where the evidence for hadronic emission is the most secure (detection of the $\pi^0$ bump discussed in section~\ref{sec:middle} \cite{piondecay}). One can note that the gamma-ray luminosity of SN 1006 is lower than for other SNRs. However, this is mainly related to its bipolar morphology as can be seen in Figure~\ref{fig:snrs}, which is a clear difference with the other TeV detected shell-type SNRs. This bipolar morphology caused by the orientation of the magnetic field in the NE-SW direction produces a reduced effective area for particle acceleration since the acceleration is more efficient when the magnetic field is parallel to the shock normal. If we correct for this effect by a renormalization factor of 0.2 \cite{berezhko2009}, the peak luminosity is comparable to other SNRs. This similar luminosity also points  towards leptonic emission. Indeed, in such scenario, the spectral shape and flux level of the gamma-ray emission is produced via the IC mechanism and only depends on the electron spectral distribution (which is similar for the five SNRs as discussed just above) and on the photon field density. Due to the omnipresence of the cosmic microwave background (CMB) and the limited effect of the infra-red photon field, one could explain why we observe such a small scatter in the gamma-ray luminosity of young shell-type SNRs.\\

However, being of hadronic or leptonic origin, the GeV-TeV gamma-ray detections of these five SNRs imply a low maximal energy for the accelerated particles of only a few $100$~TeV, well below the knee of the CR spectrum.

%%%%%%%%%%%%%%%%%%%%%%%%%%%%%%%%%%%%%%%%%%%%%%%%%%%%%%%%%
\section{Middle-age supernova remnants and the interaction with the local medium}
\label{sec:middle}

\begin{figure}[!t]
  \centering
  \includegraphics[width=12cm]{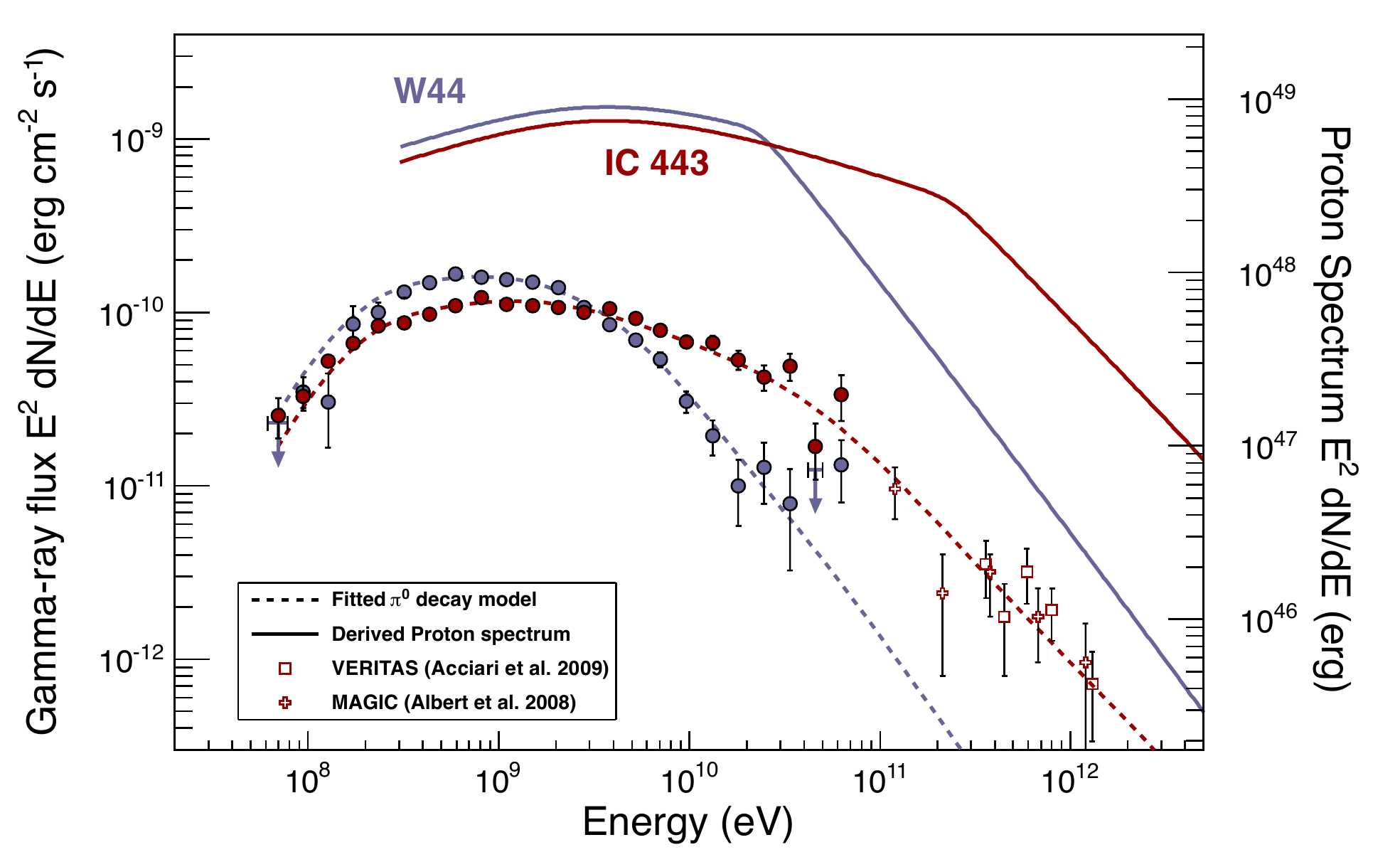}
  \caption{Proton and gamma-ray spectra determined for IC 443 and W44 from \cite{piondecay}.}
  \label{fig:bump}
\end{figure}

Excellent gamma-ray targets are provided by dense regions of the interstellar medium or in some cases dense molecular clouds interacting with middle-age SNRs ($\sim$ 10$^{11}$--10$^{12}$ s or 3,000--30,000 yr). For more information, see the recent review \cite{slane_review}. They are the most numerous subclass of GeV SNRs due to enhanced target densities for accelerated particles, which in cases of interaction with a molecular cloud can increase gamma-ray luminosities by factors of $10^2-10^3$. This enhancement has allowed the first detection of the low-energy cutoff, due to the $\pi^0$ production threshold, produced by CR proton interactions between dense clouds and SNRs IC~443 and W44 \cite{agile_w44,piondecay}. Both SNRs have underlying proton spectral indices of 2.3 with breaks near $\sim$ 100 GeV c$^{-1}$, as shown in Figure \ref{fig:bump}. While this is the first evidence for CR protons in SNRs, the acceleration process is not expected to reach very high energies for such old systems.\\

While only two SNRs show unambiguous evidence for a hadronic origin, many other middle-age SNRs appear to be likely hadronic accelerators. Puppis A is a $\sim$ 1.3$\times$10$^{11}$ s (4,000 yr) old SNR interacting with the ISM in the vicinity of a molecular cloud.  \emph{Fermi}-LAT has detected the SNR as an extended shell with a photon index of 2.1, and a break in the radio spectrum at $\sim$ 40 GHz favors a hadronic origin for simple single-zone models \cite{fermi_puppisA}. The non-detection of TeV gamma rays by H.E.S.S. indicates E$_{max} \sim 200$ GeV \cite{hess_puppisA}. In contrast, older systems such as the Cygnus Loop \cite{fermi_cygloop} and S 147 \cite{fermi_s147} show E$_{max} \sim 1$ GeV. The range in maximum energies, energy loss rate and GeV to TeV flux ratio all show weak correlation with age \cite{dermer13}, with an apparent decline in luminosity at later times, may be due to the escape of CRs from old SNRs.

\subsection{Cosmic-ray sea re-acceleration}
\label{sec:reaccel}
While middle-age SNRs are prominent gamma-ray sources, shocks in these remnants are only observed to reach velocities of a few hundred km s$^{-1}$, far below what is expected for active particle acceleration to high energies. Models have instead invoked existing CRs, compressed and possibly re-accelerated in the dense radiative filaments. Such crushed-clouds were initially proposed to explain radio emission in SNRs \cite{vanderlaan62}, and was extended to include re-acceleration of pre-existing relativistic particles \cite{1982ApJ...260..625B}. This model can explain both the bright GeV emission from proton interactions in compressed filaments, a spectral break at 1--10 GeV energies and the flat radio spectral indices that are commonly observed for SNRs interacting with molecular clouds \cite{uchiyama10}. The particular break energy may be determined by reflected shocks in the clouds \cite{Inoue2010}, Alfven wave evanescence \cite{2011NatCo...2E.194M}, or time-dependent escape from the finite-size SNR \cite{2011MNRAS.410.1577O}. An alternative scenario requires no re-acceleration, relying solely on radiative compression to explain radio and GeV spectra in old SNRs \cite{2014ApJ...784L..35T}. In these scenarios, despite the enhanced detectability of CRs, no shock acceleration to high energies is required, unlike in young SNRs.

\subsection{Cosmic-ray escape}

Another key process that studies of middle-age SNRs can help to inform is how accelerated particles are injected into the ISM to become CRs. If SNRs are the sources of Galactic CRs, then the highest energy particles have already escaped from the known middle-age SNRs, as evidenced by breaks in GeV-TeV spectra. It has long been predicted that these escaping CRs may illuminate nearby clouds. The time-dependent nature of this escape process allows the gamma-ray spectra of illuminated clouds to constrain the diffusion parameters and geometry of the system \cite{2009MNRAS.396.1629G}. Two clouds at different distances from a SNR can be expected to have different low- and high-energy spectral breaks based on the energy-dependent diffusion of CRs. SNRs W44 and W28 both have surrounding gamma-ray sources which may be interpreted as illuminated clouds, however source confusion is problematic in massive star forming regions \cite{uchiyama_w44,hess_w28,fermi_w28}. Improved resolution studies may confirm the illuminated-cloud scenario and allow the diffusion parameters to be directly determined from gamma-ray observations. Theoretical expectations are that much slower diffusion coefficients should exist within cloud regions \cite{2009ApJ...707L.179F}. Such observations may also allow tests of anisotropic diffusion, which has been suggested for dense cloud environments with large-scale ordered fields

%%%%%%%%%%%%%%%%%%%%%%%%%%%%%%%%%%%%%%%%%%%%%%%%%%%%%%%%%
\section{Pulsar wind nebulae at gamma-ray energies}
\label{sec:pwn}

\begin{figure}[!t]
\begin{center}
\includegraphics[width=0.8\linewidth]{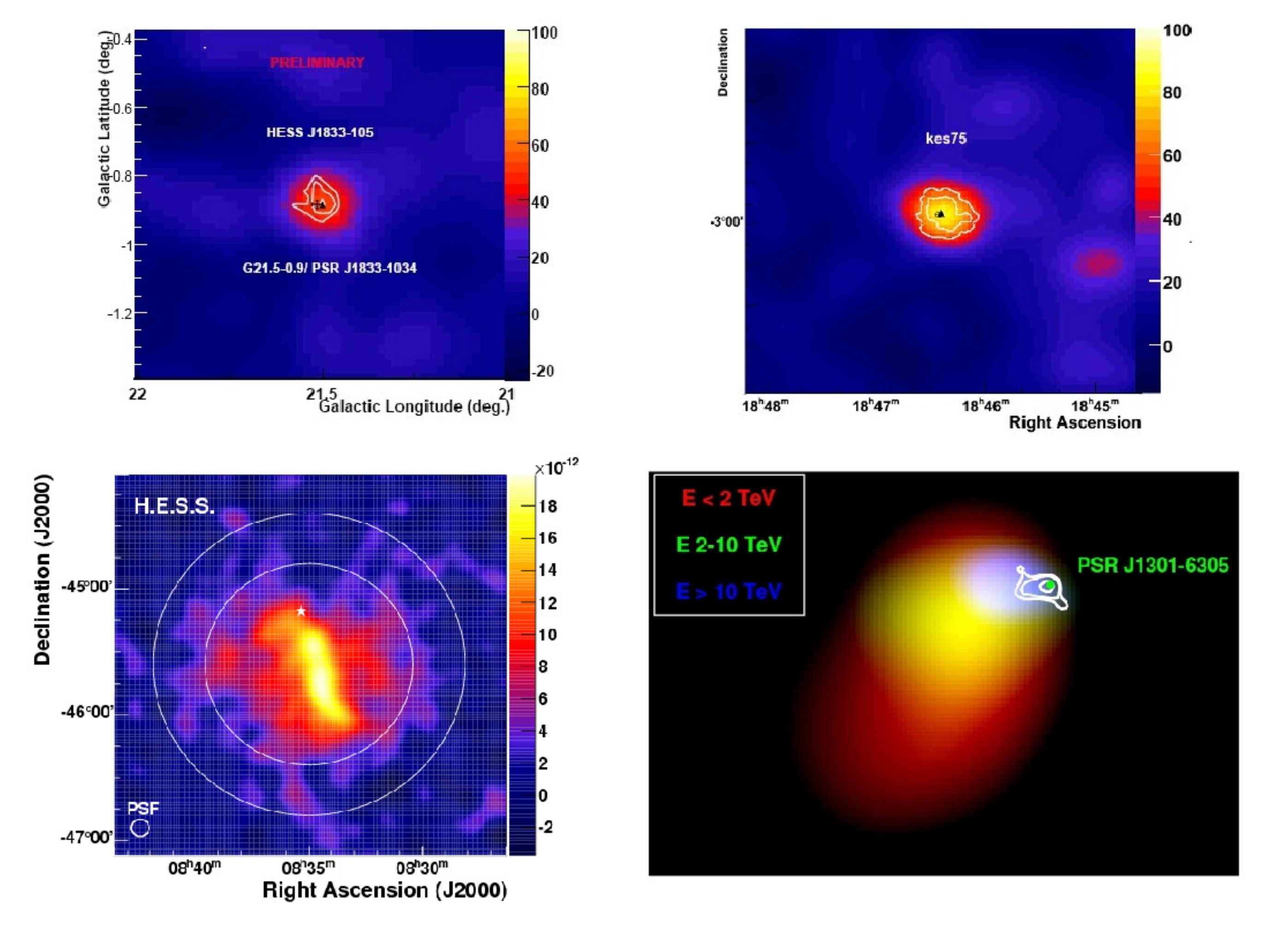}
\caption{H.E.S.S. $\gamma$-ray images four different PWNe. Top left: G21.5$-$0.9 \cite{g21} with a linear colour scale in units of counts and contour lines showing the significance contours for 4,5,6 standard deviations; Top right: Kes 75 \cite{g21} with linear colour scale in units of counts and contour lines showing significance contours for 4,5,6 standard deviations; Bottom left: Vela X \cite{velax}with linear colour scale in units of surface brightness cm$^{-2}$s$^{-1}$deg$^{-2}$; Bottom right: HESS J1303$-$631 \cite{hessj1303} shown as an energy mosaic in red for photons below 2 TeV, green for photons between 2 and 10 TeV, and blue for photons above 10 TeV. XMM-Newton X-ray contours are superimposed in white. }
\label{fig:tevpwne}
%% ==> sn1006_xmm_superposition.C 
\end{center}
\end{figure}

\subsection{A first census at gamma-ray energies}
PWNe form the most abundant class of TeV sources with more than 35 firm identifications, as of December 2014. A systematic search with the \emph{Fermi}-LAT for GeV emission in the vicinity of TeV-detected sources yielded five firmly identified high-energy gamma-ray PWNe and eleven candidates \cite{tevcat} in addition to the Crab Nebula and Vela-X which were published in separate papers. These gamma-ray PWNe are believed to be related to relatively young and energetic pulsars that power highly magnetized nebulae of a few tenths of nT to a few tens of nT (a few $\mu$G to a few hundred $\mu$G). Interestingly, the TeV luminosities do not show a significant correlation with the pulsar's spin down power \cite{klepser}. Within the context of temporal evolution of PWNe, two main classes of PWNe seem to be emerging. First, young PWNe such as the Crab Nebula (discussed in section~\ref{sec:crab}), SNR G0.9+0.1, MSH 15$-$52 and the newly discovered Crab-like VHE gamma-ray sources SNR G21.5$-$0.9 and Kes 75. Second, evolved (extended and resolved) systems (i.e. with characteristic ages $\tau>10^{11}$ s or 10$^4$ yr), such as Vela X, HESS J1825$-$137, HESS J1718$-$385 and HESS J1809$-$193. The young systems show in general a good match with the morphology seen in X-rays, while in the second group, very often the pulsar powering the TeV PWN is found offset with respect to the center of the TeV emission, with large size ratios between the X-ray and VHE gamma-ray emission regions as can be seen in Figure~\ref{fig:tevpwne}. 

%\begin{figure}[!t]
 % \centering
 % \includegraphics[width=10cm]{L_vs_edot.eps}
 % \caption{PWN GeV luminosity as a function of the pulsar spin-down power. Full markers correspond to sources with a clear PWN association at TeV energies while hollow markers correspond to sources for which the association is less clear. The black squares represent the sources detected at GeV energies, the magenta circles show the upper limits, the green pentagons are the sources showing pulsar behavior in the energy range and the blue stars represent the Crab nebula and Vela-X. PWNe detected at GeV energies are located between $10^{36}$ and $10^{39}$ erg s$^{-1}$ and with an efficiency below 10\%. From~\cite{tevcat}.}
  %\label{fig:pwn}
%\end{figure}

\subsection{Unidentified sources as potential pulsar wind nebulae candidates}
\begin{figure}[!t]
  \centering
  \includegraphics[width=10cm]{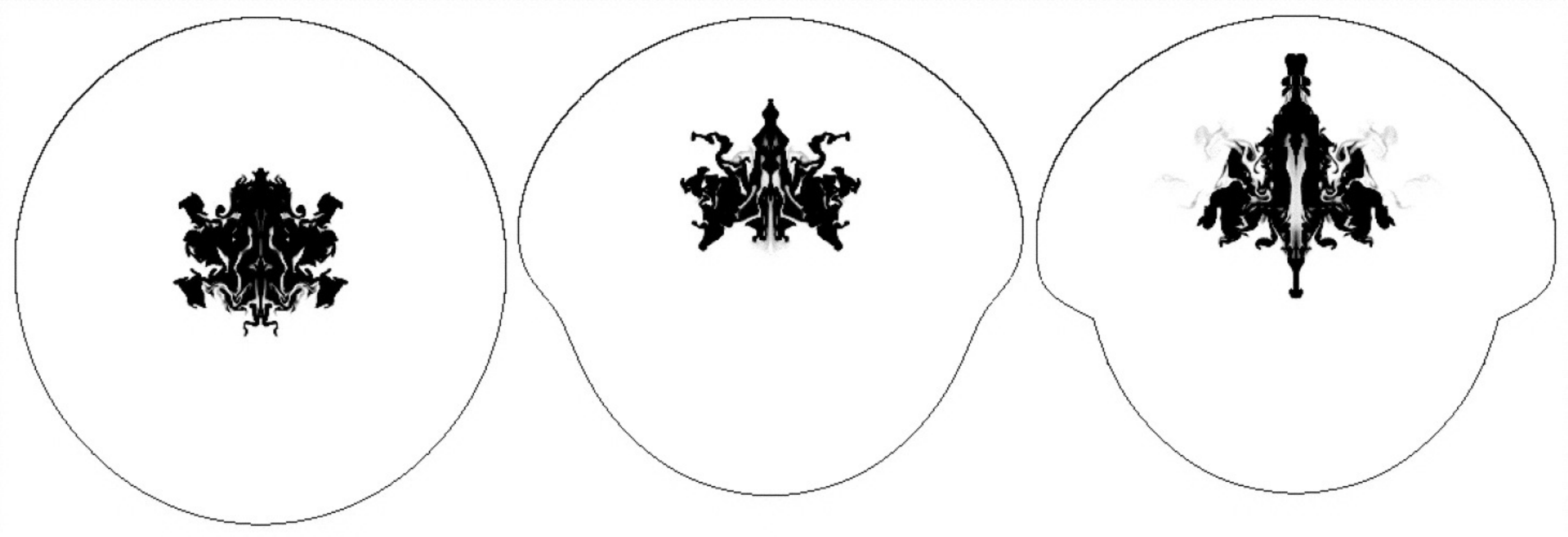}
  \caption{Simulation of the evolution of the PWN/SNR for increased density gradient scale (increased by a factor of 3 in each panel from left to right). The solid contours mark the location of the SNR shock front while the PWN is located inside these contours. From~\cite{blondin}.}
  \label{fig:blondin}
\end{figure}

The large offsets discussed above for evolved PWNe can be explained by the evolution of the SNR blast wave in an inhomogeneous medium and/or the high velocity of the pulsar, together with a low magnetic field value $\sim 0.5$ nT ($5 \mu$G). For instance, PWNe can been crushed and pushed off-center by an asymmetric reverse shock wave that resulted from the SN interaction with an asymmetric surrounding medium. This effect is clearly visible in Figure~\ref{fig:blondin} that presents simulations of the evolution of the PWN/SNR system when an increased density gradient is present in the ambient medium. In the first simulation, the asymmetry in the medium is very low, and the crushed PWN is only slightly off-center. In the two other simulations, the density gradient is increased leading to a displacement of the the crushed PWN from the center of the remnant by roughly 40\% the radius of the SNR.\\ 
The analysis of PWNe offers an excellent example of the complementarity between gamma-ray observations and the X-ray measurements that reveal the complex morphology of these sources at the arcsecond scale. Indeed, electrons emitting VHE gamma-rays are usually less energetic than X-ray-emitting ones, they do not suffer from severe radiative losses due to the relatively low magnetic field and the majority of them may survive from (and hence probe) early epochs of the PWN evolution. These electrons can produce TeV emission via IC scattering of the ambient low-energy background photons (such as CMB, diffuse Galactic infrared background, or starlight), leading to the formation of a relic PWN emitting in the VHE domain. This relic PWN is very faint or absent in X-rays because the pulsar wind electrons become too cold (due to radiative losses) and their characteristic synchrotron frequencies move outside the X-ray band. One of the biggest advantages of this scenario is that it provides a natural explanation for almost one third of the Galactic TeV sources that are still lacking a lower energy (radio and X-ray) counterpart: they form the important class of unidentified sources (UNIDs). See \cite{dejager2009} for more details. HESS J1303$-$631 was the first H.E.S.S. source classified as a UNID due to the lack of detected counterparts in radio and X-rays with Chandra \cite{mukherjee}. \cite{hessj1303} found only one plausible counterpart in the vicinity of HESS J1303$-$631: PSR J1301$-$6305 with a spin-down power of $1.70 \times 10^{29}$ W ($1.70 \times 10^{36}$ erg s$^{-1}$). The authors also presented the detection of a very weak X-ray PWN using XMM-Newton observations. This, with the energy-dependent morphology observed by H.E.S.S., led to the conclusion that HESS J1303$-$631 is an old PWN offset from the pulsar powering it. Multiwavelength observations of VHE UNIDs and dedicated pulsar searches within the extent of VHE are therefore crucial to identify PWNe systems and reveal the energetics and composition of pulsar winds.

\subsection{Pulsar wind nebulae at the origin of the positron excess}
\label{sec:positron}
Observations of PWNe are also of great importance since they could be a dominant source of leptonic CRs in the Galaxy and could explain the increase of the ratio of positrons to electrons (named positron fraction) with increasing energy as shown by PAMELA~\cite{pamelapos}, \emph{Fermi}-LAT~\cite{fermipos} and AMS-02~\cite{amspos}. Indeed, according to standard theory, the positron fraction should fall off with increasing energy, which is observed up to 10 GeV. However, above 10 GeV, an increase of the positron fraction is observed by the three experiments and could be caused by PWNe since, over time the PWN magnetic field weakens and the electrons/positrons, accelerated and trapped at the termination shock, can escape into the interstellar medium (ISM). In this context, \cite{dimauro} have used AMS-02 data to constrain pulsar/PWN origin scenarios for the positron excess, showing that this excess can be reproduced using a contribution from either the entire pulsar catalog, a few of the brightest local ($<$ 1 kpc) pulsars with ages less than 3,000 kyr, or a single local pulsar/PWN. In this latest case, the favored source of positrons is Geminga, a 340 kyr old X-ray and gamma-ray pulsar with a compact X-ray PWN, since it is the only source for which the electron emission efficiency is not required to be close to one. In the same line, \cite{yuksel} examined the implications of the \emph{Milagro} detection of extended, multi-TeV gamma-ray emission from Geminga above 20 TeV (MGRO J0634.0+1745), finding that this could reveal the existence of an ancient, powerful CR accelerator that could account for the multi-GeV positron excess. The authors also note that important constraints can be provided by gamma-ray observations, both at HE by \emph{Fermi}-LAT and VHE by ground-based detectors such as VERITAS, MAGIC or HAWC (since this source is mainly visible in the Northern Hemisphere). For instance, they have evaluated the minimal IC gamma-ray spectrum of the extended emission from Geminga explaining the \emph{Milagro} measurement at 20 TeV, showing that further study of this extended source at GeV and TeV energies would be useful to estimate the total energetics and place important constraints on the current models. Alternative solutions such as dark matter annihilation is also possible, however, a self-consistent solution in terms of purely astrophysical sources can be properly met with the help of PWNe. 

%%%%%%%%%%%%%%%%%%%%%%%%%%%%%%%%%%%%%%%%%%%%%%%%%%%%%%%%%
\section{Looking for PeVatrons}
\label{sec:pev}
\subsection{The Crab nebula: a famous electron PeVatron}
\label{sec:crab}
The Crab Nebula and its pulsar are among the best-studied objects in astronomy. It is the remnant of an historical SN, recorded in 1054 A.D., located at a distance of 2 kpc. The SN explosion left behind a pulsar, which continuously emits a wind of magnetized plasma of electron/positron pairs \cite{pulsar}. These particles lose energy by synchrotron radiation, visible from radio up to hundreds of MeV, and IC scattering of the generated synchrotron and ambient radiation fields, detected at gamma-ray energies. The large-scale integrated emission from the Crab Nebula is expected to be steady within a few percent and is thus often used to cross-calibrate X-ray and gamma-ray telescopes and to check their stability over time. Recently, variability in the X-ray flux from the nebula by $\sim$3.5\% per year has been detected (1999 -- 2008), setting limits on the accuracy of this practice \cite{wilson2011}. In addition, instabilities in the flux of high-energy gamma rays have been reported in recent years by AGILE and the \emph{Fermi}-LAT (see \cite{buehler} for a review). These flares have all shown increased emission (up to a factor 20) from the synchrotron component of the Crab Nebula while emission from the IC component of the nebula as well as the Crab pulsar remained consistent with the average level. An approximate flux doubling timescale of 6 hr was reported for the two brightest ones. The detection of synchrotron photons up to energies of  $>1$ GeV confirms that electrons are fastly accelerated to energies above 1 PeV in the Crab Nebula, with a magnetic field value as high as 100 nT (1 mG). These brief time scales and the requirement that the emission volume be causally connected imply that the flaring region must be compact ($< 10^{-2}$ pc) and the acceleration be extremely efficient (which poses serious challenges on scenarios invoking DSA). Structures this small are found only in the inner part of the nebula, close to the termination shock. Several new ideas have been proposed to explain these recurring flares ($\sim$1 per year), one of them being magnetic reconnection \cite{dsa,cerutti,sironi}, but they still remain mysterious and, to date, despite extensive efforts a detection of the flares outside of the HE gamma-ray band remains elusive. This absence of plausible counterparts at other wavelengths is certainly one of the most surprising aspects of the flare phenomenon and reinforce very well the needs of future gamma-ray observations with higher sensitivity and improved angular resolution. 

\begin{figure}[!t]
  \centering
  \includegraphics[width=10cm]{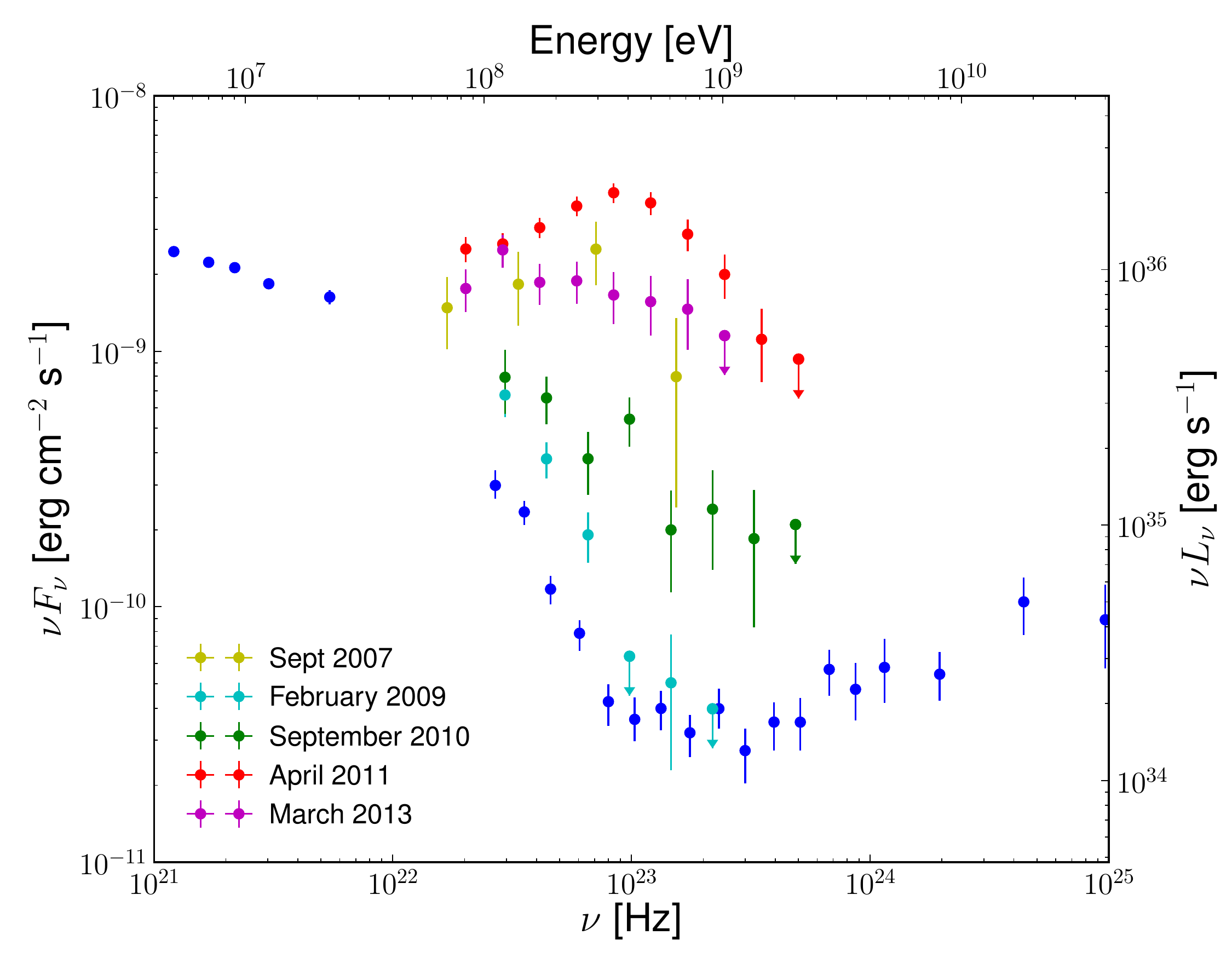}
  \caption{Spectral energy distribution at the maximum flux level for the five Crab nebula flares (from~\cite{buehler}). The blue points represent the average Crab nebula flux values. The spectral points from September 2007 were obtained by AGILE, while the other points were obtained with the \emph{Fermi}-LAT.}
  \label{fig:crab}
\end{figure}

\subsection{Evolution of the cosmic-ray acceleration efficiency : reaching the knee with supernova remnant shocks}
The recent GeV and TeV detections of SNRs confirm the theoretical predictions that SNRs can operate as powerful CR accelerators. However, if these objects are responsible for the bulk of galactic CRs, they should be able to accelerate protons and nuclei at least up to $10^{15}$~eV and therefore act as hadronic PeVatrons. As discussed in this review, even if $\pi^0$ decay would be the correct
model to interpret the gamma-ray data, none of the remnants observed so far show evidence for protons with energies up to $3 \times 10^{15}$ eV.
 In this respect, Tycho could be considered as a half-PeVatron at least, since there is no evidence of a cut-off in the VERITAS data. There is, however indirect evidence for protons with these high energies in SNRs, as shown by the stripes of non-thermal X-ray emission on the surface of the historical SNR Tycho \cite{bykov}. It may thus be that the spectrum of non thermal particles extends to PeV energies only during a relatively short period of the evolution of the remnant since high energy particles are the first to escape from the SNR shock \cite{gabici}. For this reason one may expect spectra of secondary gamma rays extending to energies beyond 10 TeV only from less than 1~kyr old SNRs. One may wonder how many PeVatrons are expected to be detectable in our Galaxy. A simple estimate has been provided by \cite{gabici}: assuming a rate of $\sim$3 SNe per century in our Galaxy, this directly implies that only a dozen of PeVatrons are present in the Galaxy on average and hence that they are likely to be distant and weak. 

%%%%%%%%%%%%%%%%%%%%%%%%%%%%%%%%%%%%%%%%%%%%%%%%%%%%%%%%%
\section{Outlook}
\label{sec:discussion}
These results emphasize the importance of TeV observations by the future generation of Cherenkov telescopes such as the Cherenkov Telescope Array (CTA) \cite{cta} which will have a better effective area in the energy range already covered but that will also allow the observation of sources such as Tycho up to energies higher than 100 TeV, therefore constraining the maximal energy at which protons are being accelerated in young SNRs. Monte-Carlo simulations of shell-type SNRs have already been carried out for different CTA array layouts, assuming a uniform exposure time of 20 hr everywhere along the Galactic Plane. In this purpose, the morphological and spectral characteristics of three SNRs (RX~J1713.7-3946, Vela Junior and RCW~86), as measured with H.E.S.S., together with their respective distance estimates, have been used to simulate sources throughout the inner Galaxy (\cite{cta_gabici}, \cite{cta_renaud}). This leads to $\sim$20--70 detectable TeV SNRs, among which $\sim$7--15 would be resolved with CTA with a configuration optimized for providing the best sensitivity over the whole energy range. It is striking to note that, thanks to its increased sensitivity, CTA will have the capability to detect SNRs as luminous as RX~J1713.7$-$3946, Vela Junior, or RCW 86 up to the other side of the Galaxy. A similar study was performed by \cite{cta_emma} showing that $\sim 300$ PWNe should be detected by CTA. We can therefore expect that these future gamma-ray observations will provide a complete census of the CR accelerators in our Galaxy.
%%%%%%%%%%%%%%%%%%%%%%%%%%%%%%%%%%%%%%%%%%%%%%%%%%%%%%%%%

%%%%%%%%%%%%%%%%%%%%%%%%%%%%%%%%%%%%%%%%%%%%%%%%%%%%%%%%%
\section*{Acknowledgements}
MLG and JWH thank F. Acero and all the members of the \emph{Fermi}-LAT and H.E.S.S. GALACTIC working groups for valuable discussions. 
%%%%%%%%%%%%%%%%%%%%%%%%%%%%%%%%%%%%%%%%%%%%%%%%%%%%%%%%%

%%%%%%%%%%%%%%%%%%%%%%%%%%%%%%%%%%%%%%%%%%%%%%%%%%%%%%%%%


\begin{thebibliography}{00}
%%%%%%%%%%%%%%%%%%%%%%%%%%%%%%%%%%%%%%%%%%%%%%%%%%%%%%%%%
%   the references should be in order of citation in the text

%
% introduction
%
\bibitem{dsa} Lemoine, M. \& Pelletier, G., this volume
\bibitem{Bell1987} Bell, A.~R.\ , MNRAS 225 (1987) 615 
\bibitem{Jones1991} Jones, F.~C., \& Ellison, D.~C.\ , SSRv 58 (1991) 259 
\bibitem{Axford1977} Axford, W.~I., Leer, E., \& Skadron, G.\ , International Cosmic Ray Conference, 11 (1977) 132 
\bibitem{Krymskii1977} Krymskii, G.~F.\ , Akademiia Nauk SSSR Doklady 234 (1977) 1306 
\bibitem{Bell1978} Bell, A.~R.\ , MNRAS 182 (1978) 147 
\bibitem{BO1978} Blandford, R.~D., \& Ostriker, J.~P.\ , ApJL 221 (1978) L29 
\bibitem{Berezhko1999} Berezhko, E.~G., \& Ellison, D.~C.\, ApJ 526 (1999) 385 
\bibitem{Woltjer1970} Woltjer, L.\ , Interstellar Gas Dynamics 39 (1970) 229 
\bibitem{Cioffi1988} Cioffi, D.~F., McKee, C.~F., \& Bertschinger, E.\ , ApJ 334 (1988) 252 
\bibitem{Drury1989} Drury, L.~O., Markiewicz, W.~J., \& Voelk, H.~J.\ , A\&A 225 (1989) 179 
\bibitem{Dwarkadas1998} Dwarkadas, V.~V., \& Chevalier, R.~A.\ , ApJ 497 (1998) 807 
\bibitem{Dwarkadas2013} Dwarkadas, V.~V.\ , MNRAS 434 (2013) 3368 
\bibitem{Chevalier1999} Chevalier, R.~A.\ , ApJ 511 (1999) 798 
\bibitem{Inoue2012} Inoue, T., Yamazaki, R., Inutsuka, S.-i., \& Fukui, Y.\ , ApJ 744 (2012) 71 
\bibitem{pwn_review} Gaensler, B.~M., \& Slane, P.~O.\ , ARA\&A 44 (2006) 17 
\bibitem{pulsar} Grenier, I. \& Harding, A., this volume
\bibitem{Gelfand2009} Gelfand, J.~D., Slane, P.~O., \& Zhang, W.\ , ApJ 703 (2009) 2051 
\bibitem{Martin2012} Mart{\'{\i}}n, J., Torres, D.~F., \& Rea, N.\ , MNRAS 427 (2012) 415 
%
% section 2
%
\bibitem{2014ApJ...782L...2I} Indebetouw, R., Matsuura, M., Dwek, E., et al.\ , ApJL 782 (2014) LL2 
\bibitem{Ellison2000} Ellison, D.~C., Berezhko, E.~G. \& Baring, M.~G.\ , ApJ 540 (2000) 292 
\bibitem{Reville2013} Reville, B. \& Bell, A.~R.\ , MNRAS 430 (2013) 2873 
\bibitem{2014ApJ...796...82Z} Zanardo, G., Staveley-Smith, L., Indebetouw, R., et al.\ , ApJ 796 (2014) 82 
\bibitem{hess_lmc} Abramowski, A., et al., Science 347 (2015)  406
\bibitem{dens_sn1987a} Mattila, S., et al., ApJ 717 (2010) 1140
%\bibitem{Berezhko2011} Berezhko, E.~G., Ksenofontov, L.~T., Voelk, H.~J.\ , ApJ, 732 (2011) 58 
\bibitem{fermi_casa}{Abdo, A. A., et al.}, ApJL 710 (2010) L92
\bibitem{magic_casa}{Albert, J., et al.}, A\&A 474 (2007) 937
\bibitem{veritas_casa}{Acciari, V. A., et al.}, ApJ 714 (2010) 163 
\bibitem{fermi_tycho}{Giordano, F., et al.}, ApJL 744 (2012) L2 
\bibitem{veritas_tycho}{Acciari, V. A., et al.}, ApJL 730 (2011) L20
\bibitem{yuan_casA} Yuan, Y., Funk, S., J{\'o}hannesson, G., et al.\ , ApJ 779 (2013) 117 
\bibitem{morlino}{ Morlino, G. \& Caprioli, D.}, A\&A 538 (2012) A81
\bibitem{chiotellis12} Chiotellis, A., Schure, K.~M., \& Vink, J.\ , A\&A 537 (2012) A139 
\bibitem{hess_kepler} Aharonian, F., Akhperjanian, A.~G., Barres de Almeida, U., et al.\ , A\&A 488 (2008) 219 
\bibitem{patnaude12} Patnaude, D.~J., Badenes, C., Park, S., \& Laming, J.~M.\ , ApJ 756 (2012) 6 
\bibitem{2014SerAJ.189...41D} De Horta, A.~Y.,  Filipovic, M.~D., Crawford, E.~J., et al.\, Serbian Astronomical  Journal 189 (2014) 41 
\bibitem{2015ApJ...798...98Z} Zoglauer, A., Reynolds, S.~P., An, H., et al.\, ApJ 798 (2015) 98 
\bibitem{hess_g1.9} H.E.S.S.~Collaboration, Abramowski, A., Aharonian, F., et al.\,  MNRAS 441 (2014) 790 
\bibitem{cta_young} Acharya, B.~S., Aramo, C., Babic, A., et al.\, Astroparticle Physics 62 (2015) 152 
%
% section 3: TeV shell SNRs
%
\bibitem{aharonian_rxj1} Aharonian, F., et al., Nature 432 (2004) 75
\bibitem{aharonian_rxj2} Aharonian, F., et al., A\&A 437 (2005) L7
\bibitem{aharonian_velajr1} Aharonian, F., et al., A\&A 464 (2011) 235
\bibitem{aharonian_rcw86} Aharonian, F., et al., ApJ 692 (2011) 1500
\bibitem{acero_sn1006} Acero, F., et al., A\&A 512 (2010) A62
\bibitem{acero_1731} Abramowski, A., et al., A\&A 531 (2011) A81
\bibitem{fermi_rxj} Abdo, A.~A., et al., ApJ 734 (2011) 28
\bibitem{fermi_velajr} Tanaka, T., et al., ApJL 740 (2011) L51     
\bibitem{yuan_rcw86} Yuan, Q., Huang, X., Liu, S., \& Zhang, B., ApJ 785 (2014) L22
\bibitem{wang_rxj} Wang, Z. R., Qu, Q.-Y. \& Chen, Y., A\&A 318 (1997) L59 
\bibitem{rxj_rosat} Pfeffermann, E. \& Aschenbach, B., Proc. Roentgenstrahlung from the Universe ed. H. U. Zimmermann, J. H. Tr\"umper, \& H. Yorke, (1996) 267
\bibitem{UchiyamaAsca} Uchiyama, Y., Takahashi, T. \& Aharonian, F. A., PASJ 54 (2002) L73
\bibitem{berezhko}{ Berezhko, E. G. \& Voelk, H. J.}, A\&A 511 (2010) A34
\bibitem{federici} Federici, S., Pohl, M., Telezhinsky, I., Wilhelm, A. \& Dwarkadas, V.V., A\&A (2015) 577, 12
\bibitem{Acero2013} Acero, F., Gallant, Y., Ballet, J., Renaud, M. \& Terrier, R., A\&A 551 (2013) A7 
\bibitem{piondecay} Ackermann, M., et al., Science 339 (2013) 807
\bibitem{berezhko2009} Berezhko, E. G., Ksenofontov, L. T. \& Voelk, H. J., A\&A 505 (2009) 169
\bibitem{acero_shell} Acero, F., et al., A\&A (2015) submitted
%
% section 4: middle aged SNRs 
%
\bibitem{slane_review} Slane, P., Bykov, A., Ellison, D.~C., Dubner, G., \& Castro, D.\, SSRv (2014) 26 
\bibitem{agile_w44} Giuliani, A., Cardillo, M., Tavani, M., et al.\, ApJL 742 (2011) L30 
\bibitem{fermi_puppisA} Hewitt, J.~W., Grondin, M.-H., Lemoine-Goumard, M., et al.\, ApJ 759 (2012) 89 
\bibitem{hess_puppisA} H.~E.~S.~S.~Collaboration, :, Abramowski, A., et al.\ 2014, arXiv:1412.6997 
\bibitem{fermi_cygloop} Katagiri, H., Tibaldo, L., Ballet, J., et al.\, ApJ 741 (2011) 44 
\bibitem{fermi_s147} Katsuta, J., Uchiyama, Y., Tanaka, T., et al.\, ApJ 752 (2012) 135 
\bibitem{dermer13} Dermer, C.~D., \& Powale, G.\, A\&A 553 (2013) A34 
\bibitem{vanderlaan62} van der Laan, H.\, MNRAS 124 (1962) 179 
\bibitem{1982ApJ...260..625B} Blandford, R.~D., \& Cowie, L.~L.\, ApJ 260 (1982) 625 
\bibitem{uchiyama10} Uchiyama, Y., Blandford, R.~D., Funk, S., Tajima, H., \& Tanaka, T.\, ApJL 723 (2010) L122 
\bibitem{Inoue2010} Inoue, T., Yamazaki, R., \& Inutsuka, S.-i.\, ApJL 723 (2010) L108 
\bibitem{2011NatCo...2E.194M} Malkov, M.~A., Diamond, P.~H., \& Sagdeev, R.~Z.\, Nature Communications 2 (2011) 194 
\bibitem{2011MNRAS.410.1577O} Ohira, Y., Murase, K., \& Yamazaki, R.\, MNRAS 410 (2011) 1577 
\bibitem{2014ApJ...784L..35T} Tang, X., \& Chevalier, R.~A.\, ApJL 784 (2014) L35 
\bibitem{2009MNRAS.396.1629G} Gabici, S., Aharonian, F.~A., \& Casanova, S.\, MNRAS 396 (2009) 1629 
\bibitem{uchiyama_w44} Uchiyama, Y., Funk, S., Katagiri, H., et al.\, ApJL 749 (2012) L35 
\bibitem{hess_w28} Aharonian, F., Akhperjanian, A.~G., Bazer-Bachi, A.~R., et al.\, A\&A 481 (2008) 401 
\bibitem{fermi_w28} Hanabata, Y., Katagiri, H., Hewitt, J.~W., et al.\, ApJ 786 (2014) 145 
\bibitem{2009ApJ...707L.179F} Fujita, Y., Ohira, Y., Tanaka, S. J., \& Takahara, F. \ , ApJL 707 (2009) L179 
%
% section 5: PWNe
%
\bibitem{tevcat} Acero, F., Ackermann, M., Ajello, M., et al., ApJ 773 (2013) 77
\bibitem{klepser} Klepser, S., et al., Proceedings of the 33rd ICRC (2013)
\bibitem{g21} Djannati-Atai, A., et al., Proceedings of the 30th ICRC (2008) vol. 2 p823-826 
\bibitem{velax} Abramowski, A., et al., A\&A, 548 (2012) A38 
\bibitem{blondin} Blondin, J. M., Chevalier, R. A., Frierson, D. M. \ ApJ 563 (2001) 806
\bibitem{dejager2009} de Jager, O. C. \& Djannati-Atai, A., in Neutron Stars and Pulsars, Astrophysics and Space Science Library, 2009, Vol. 357 (Berlin: Springer), 451
\bibitem{mukherjee} Mukherjee, R. \& Halpern, J. P., ApJ 629 (2005) 1017
\bibitem{hessj1303} H.E.S.S. Collaboration, Abramowski, A., Acero, F., et al., A\&A 548 (2012b) A46
\bibitem{pamelapos} Adriani, O., Barbarino, G. C., Bazilevskaya, G. A., et al., Nature 458 (2009) 607
\bibitem{fermipos} Abdo, A. A., Ackermann, M., Ajello, M., et al., Phys. Rev. Lett. 102 (2009) 18
\bibitem{amspos} Aguilar, M., Alberti, G., Alpat, B., et al., Phys. Rev. Lett. 110 (2013) 141102
\bibitem{dimauro} Di Mauro, M., Donato, F., Fornengo, N., et al., JCAP 1404 (2014) 006
\bibitem{yuksel} Yuksel, H., Kistler, M. D., Stanev, T., Phys. Rev. Lett. 103 (2009) 5
%
% section 6: PeV CRs
%
\bibitem{wilson2011} Wilson-Hodge, C. A., et al., ApJL 727 (2011) L40
\bibitem{buehler} Buehler, R. \& Blandford, R., Reports on Progress in Physics 77 (2014) 6
\bibitem{cerutti} Cerutti, B., Uzdensky, D. A. \& Begelman, M. C., ApJ 746 (2012) 148
\bibitem{sironi} Sironi, L. \& Spitkovsky, A., ApJ 741 (2011) 39
\bibitem{bykov} Bykov, A. M., Ellison, D. C., Osipov, S. M., et al., ApJ 735 (2011) L40
\bibitem{gabici} Gabici, S. \& Aharonian, F. A., ApJ 665 (2007) L131
\bibitem{cta} Knoedlseder, J., to be published in a special issue of C.R. Physique, continuation of the present one (2016)
\bibitem{cta_gabici} Acero, F.,  et al., Astroparticle Physics 43 (2013) 276
\bibitem{cta_renaud} Renaud, M., et al. 2011, CRISM proceedings, Memorie della Societa Astronomica Italiana
\bibitem{cta_emma} de Ona Wilhelmi, E., et al., Astroparticle Physics 43 (2013) 287

\end{thebibliography}
\end{document}